# A reduction-to-absurdity approach using absolutely smooth solid surfaces to unveil the origins of wetting

Qiao Liu, Jiapeng Yu, Hao Wang*

The Laboratory of Heat and Mass Transport at Micro-Nano Scale, College of Engineering, Peking University, Beijing 100871, People's Republic of China

___

**ABSTRACT**

Contact angle hysteresis and generation of dynamics angle are two fundamental phenomena about the contact angle deviation from the equilibrium state. Roughness on the solid surface, disjoining pressure in the thin film, and liquid-solid adhesion have all been considered as the origins of the phenomena. This work for the first time made a reduction to absurdity by employing absolutely smooth solid surfaces in ultra-large-scale molecular dynamic simulation. The results showed that the equilibrium angles were well established on the absolutely smooth surface just as regular, while the hysteresis and the dynamic angle vanished. The critical structure of the convex nanobending for advancing contact lines vanished as well. In contrast, the solids that made of atoms, even at the minimum roughness, would bring significant angle deviation and convex nanobending. A 3D observation was further made using state-of-the-art helium ion microscopy for the first time revealing the ubiquitous nanoscopic distortion along the contact line on atomically smooth surfaces. The results answer the question of the origin of the angle deviation, that the hysteresis and dynamic angle can be unified to originate from the friction, either static or dynamic; and at the same time resolve the puzzle of the hysteresis existence on reported smooth surfaces, showing that even the minimum i.e. atomic roughness is great enough to be effective on the contact line. The results are against the disjoining pressure theory which ignored the roughness and predicted the occurrence of the hysteresis on absolutely smooth solids.

___

1. Introduction

Wetting plays important roles in large-scale systems such as deposition, coating, and oil recover,[1] and has great impacts on small-scale flows therefore important to emerging technologies such as electronics cooling,[2-4] micro and nanofluidics,[5-6] self-assembly,[7-8] friction,[9] and various biological processes.[10-12] The contact angle hysteresis and the generation of dynamics angle during wetting are two basic phenomena remaining controversial for many years.

Liquids show contact angle hysteresis on solids. They will not advance if the contact angle is smaller than a critical value called advancing static contact angle, and will not recede if the contact angle is bigger than a critical value called receding static contact angle. The difference between the advancing and the receding static contact angle is defined as the contact angle hysteresis.[13-17] The solid roughness has long been considered as an important factor for the hysteresis, but still full of controversy: First is about the exact role of the roughness. Does it create or just enhance the hysteresis? There is no doubt about the enhancing effect. Ample experiments like those of Johnson and Dettre[18] and Tammer et al.[19] have concluded that contact angle hysteresis increased with the roughness. However, still there is no clear evidence showing that the hysteresis is rooting from the roughness. Some studies attributed the hysteresis to the disjoining pressure in the thin liquid film near the contact line.[20] On absolutely smooth homogeneous solid surfaces, Starov et al.[21] and Arjmandi-Tash et al.[22] have shown that the hysteresis can be calculated via disjoining/conjoining pressure isotherms.

A related controversy is, what is the effective roughness scale for the hysteresis? A few of theoretical studies assumed that the surface with root-mean-square (RMS) roughness < 100 nm can be treated as smooth enough to eliminate the contact angle hysteresis.[23] However, Tammer et al.[19], Lam et al.[24], Extrand et al.[25] and Miller et al.[26-27] have found non-negligible hysteresis on silicon wafers, polymer, and polyamide films whose RMS roughness were far below 100 nm. The hysteresis was even observed for liquids on substrates coated by a thin liquid film.[28] These facts on seemingly ideally smooth surface has been the motivation of using the disjoining pressure to explain the hysteresis. As stated by Arjmandi-Tash et al.,[22] "The phenomenon of contact angle hysteresis is usually attributed to the roughness and/or chemical heterogeneity of the surface. Although these properties of the substrate play a significant role in the contact angle hysteresis, they are not the sole reason for the hysteresis phenomenon. Convincing proof of the existence of contact angle hysteresis even on smooth homogeneous surfaces has been presented earlier."[25, 27, 29-30]

Contact line starts to move when the external force on the liquid is great enough to make the angle beyond the hysteresis. The dynamic contact angle is then generated varying with the spreading speed. One focus of the long-standing debate is about the energy dissipation channel. The hydrodynamic models[31-33] assume negligible friction dissipation at the contact line thus the microscopic contact angle at the contact line, is a constant equal to the equilibrium angle. While the molecular kinematic theory (MKT) [34-36] considers the importance of local dissipation at the moving contact line. The microscopic contact angle is thus dependent on the moving speed. Recently Chen *et al.* [37] made a pioneering study which clarified the nanoscopic morphology near the advancing contact line using an AFM. The key feature is a shoe-tip-like convex nanobending region within 20 nm of the substrate. The microscopic contact angle extracted at the root of the nanobending was found significantly dependent on the moving speed. On the other hand, a few of studies have discussed the effect of surface roughness. Semal *et al.* [38] analyzed the dynamic contact angle of a sessile drop spreading spontaneously on Langmuir−Blodgett multilayer substrates which allowed the roughness to be adjusted. The dynamic angles were larger with the greater roughness. But again similar to the debates on the contact angle hysteresis, it is still not settled about the exact role of the roughness in dynamic wetting. Does it create the dynamic angle variation or just an enhancing factor? And also what is the minimum efficient scale, does atomic roughness smooth enough to eliminate the dynamic angle variation? The debate held. A most recent simulation work[39] demonstrated that even nanometer-sized surface defects can produce a measurable effect on advancing and receding liquid fronts.

The puzzles root from the multi-factor and multi-scale nature of the phenomena. If we make an analysis of major competing factors in the system, they are limited to three components: a. the cohesion forces within the liquid; b. the adhesion forces between the liquid and solid whose direction is normal to the solid surface; and c. friction whose direction is lateral respect to the solid surface. Many studies have analyzed the influences of different components, but decisive conclusions are still hard to make since the investigations always involve the whole system. In the current study, an absolutely smooth surface was employed for the first time to decouple the tangled factors. State-of-the-art atomic force microscopy and helium ion microscopy were employed to achieve nanoscale detection at the contact line. The simulation and experimental results give clear answers to the questions about the exact role of the roughness and the origins of the wetting behaviors.

## 2. Simulation and experimental methods

The system of the ultra-large-scale molecular dynamic simulation consisted of a solid substrate and a liquid droplet with diameter up to 50 nm. The droplets in our simulations were made of argon atoms with each mass $m_{ll}$ =39.95 g•mol$^{-1}$. We adopted the Lennard-Jones 9/3 wall [40] as the absolutely smooth surface and refer to it as surface A. It was a mathematically ideal wall without any defects or crystal gap. Surface A interacted with liquid atoms by generating a force in the direction perpendicular to the surface, the interaction was the average by integrating interactions between every liquid atom and whole substrate A. Therefore, there was no friction on the surface A. The interaction energy between the droplet atoms and surface A can be denoted by

$$\varphi_{ab} = \varepsilon_{ab} \left[ \frac{2}{15} \left( \frac{\sigma_{ab}}{r} \right)^9 - \left( \frac{\sigma_{ab}}{r} \right)^3 \right] \quad (1)$$

where $\varepsilon$ and $\sigma$ were the usual Lennard-Jones parameters, determining the strength and the size of the atom as it interacted with the wall. Subscript $a$ and $b$ denote different kinds of atoms, which can be replaced by $l$ (liquid) and $s$ (solid) respectively.

We conducted two kinds of regular surfaces i.e. surface B and surface C using platinum atoms with each mass $m_{ss}$ =195.11 g•mol$^{-1}$. Surface B was atomically smooth single crystal with FCC structure with lattice constant 0.392 nm. Surface C was with roughness elements whose height and gap were both 1.0 nm. The interaction energy between droplets and real surface B and C was the standard Lennard-Jones potential[41] i.e.

$$\varphi_{ab} = 4\varepsilon_{ab} \left[ \left( \frac{\sigma_{ab}}{r} \right)^{12} - \left( \frac{\sigma_{ab}}{r} \right)^6 \right] \quad (2)$$

For liquid-liquid interaction, $\varepsilon_{ll}$ and $\sigma_{ll}$ were 0.2404 kcal•mol$^{-1}$ and 0.3405 nm respectively. For solid-solid interaction, $\varepsilon_{ss}$ and $\sigma_{ss}$ are 12.0181 kcal•mol$^{-1}$ and 0.2475 nm respectively[41]. For liquid-solid interaction, we tested a series of values. Detailed parameter settings refer to the Supporting Information.

One of the experimental systems was based on a state-of-the-art helium ion microscopy (HIM, ZEISS ORION NanoFab, Carl Zeiss Microscopy). It has sub-nanometer higher resolution imaging of uncoated biological samples and higher surface sensitivity than SEM,[42-43] and importantly for the current study, it has enough depth of focus allowing a 3D imaging for the contact line region. Ionic liquid 1-Butyl-3-methylimidazolium tetrafluoroborate was selected as the working liquid since it kept non-volatile in the vacuum chamber of the microscopy. Four kinds of solid substrates with different surface roughness were employed i.e. mica, silicon wafer, copper plate, and coarse copper plate.

The other experimental system was based on tapping-mode atomic force microscopy (AFM, MFP-3D-BIO, Asylum Research), which had the low-noise performance for high-resolution imaging of the most delicate samples like gel, liquid surfaces or proteins. [37, 44-45] The experimental material included PEG400, PPG2000 and glycerol droplets on silicon wafer ((100)-oriented, p-doped, 0.05~0.2 Ω•m, Micro-Nano Machining Center, Peking University) and quartz (Reagent Management Platform, Peking University), which have different scales of surface roughness. The dynamic wetting process was achieved by depositing millimeter droplets on the solid substrates. The thin film profiles of the advancing contact line were scanned and error analysis of the measurement was identical to Chen *et al.*.[37]

**3. Results and discussions**

3.1. Contact angle hysteresis

After the equilibrium state of the droplet on solid being reached, we imposed a lateral body force on the droplets to investigate the contact angle hysteresis. The lateral force, $2 \times 10^{-15}$ N, pushed and might change the droplet morphology. The maximum difference between the advancing and receding contact static angles was recorded to represent the hysteresis.[46] As shown in Fig. 1, with the same equilibrium contact angle, 56°, atomic surface B had significant hysteresis while the absolutely smooth surface A had no hysteresis at all. The droplet simply slid on surface A under the lateral force without shape change. The wettability didn't change the trend, that is, the hysteresis on A was always zero despite the equilibrium angle variation, as shown in Table 1. For atomically smooth surface B, the hysteresis increased with the roughness as well as the wettability as shown in Table 1.

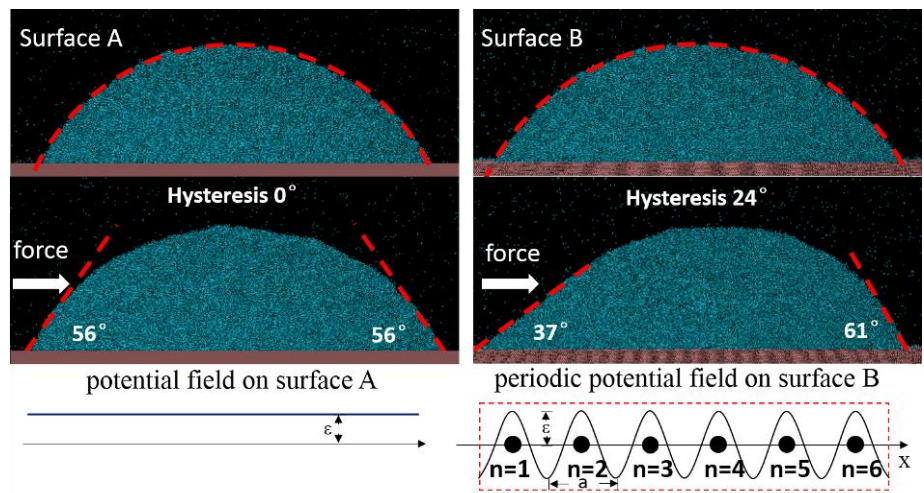

Figure 1. The snapshots of droplets at equilibrium states (top row) and hysteresis (bottom row) on different surfaces. Surface A was absolutely smooth and no hysteresis occurred. Surface B was made of atoms and atomically smooth.

Table 1. Simulated contact angle hysteresis and interfacial static friction for different systems. $\theta_0$ is the equilibrium contact angle. Static friction was calculated based on Eq. 4.

| System | Receding angle | Advancing angle | Hysteresis | Static friction(kcal·(mol·Å)$^{-1}$ |
|---|---|---|---|---|
| $\theta_0$=88°,surface A | 88° | 88° | 0° | 0 |
| $\theta_0$=56°,surface A | 56° | 56° | 0° | 0 |
| $\theta_0$=30°,surface A | 30° | 30° | 0° | 0 |
| $\theta_0$=56°,surface B | 37° | 61° | 24° | 17.34 |
| $\theta_0$=56°,surface C | 32° | 70° | 38° | 28.91 |
| $\theta_0$=97°,surface B | 84° | 103° | 19° | 10.52 |
| $\theta_0$=40°,surface B | 20° | 71° | 51° | 41.62 |

The comparison between the absolutely smooth and atomically smooth surfaces gave clear evidence that the hysteresis was rooting from the static friction on the surface. The static friction was occurring when the liquid molecules came to the lateral energy barriers on the surface. The sharp contrast on the absolutely smooth and atomically smooth surfaces also indicated that even atomic-scale roughness is effective enough for the hysteresis.

A straightforward experimental observation was conducted to further confirm the simulation conclusion. As shown in Fig. 2, using state-of-the-art helium ion microscopy we revealed the 3D morphology near the static contact line. The solids included mica which was atomically smooth with RMS roughness as low as 0.12 nm according to the AFM scanning. The contact line appeared smooth when observing at the microscale, but numerous nano distortions and wrinkles were observed when coming to nanoscale as shown in Fig. 2. The distortions and wrinkles were everywhere along the contact line, which was strong evidence that the atomic smoothness was not smooth enough for the contact line. The distortions and wrinkles would get more significant when the substrate roughness was increased. The statistical information of the contact line distortion on different surfaces was measured as shown in the Supporting Information.

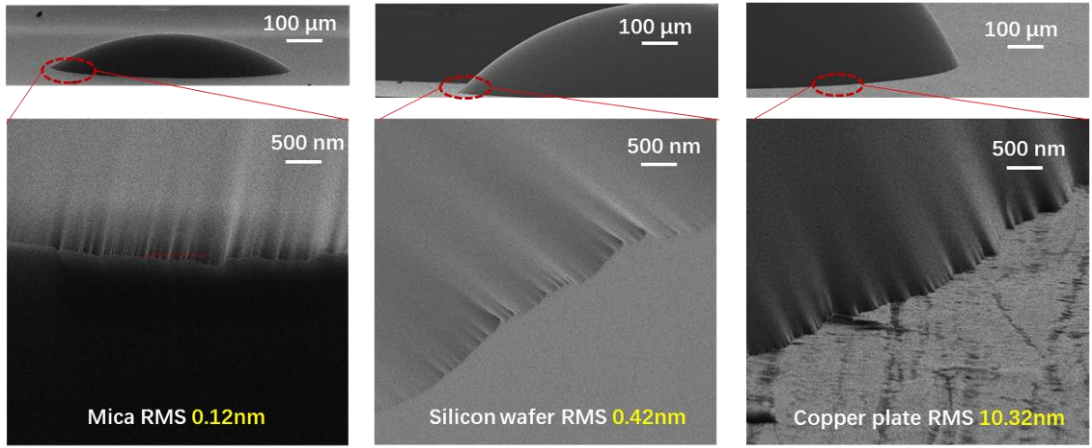

Fig. 2. Helium ion microscopy for the first time revealed the 3D nanoscopic morphology near the static contact line. The substrate distorted the contact line into numerous tiny subsections, even though the contact line appeared smooth in microscale observation. The distortion then produced numerous wrinkles on the vicinity liquid film. The atomic smoothness like mica surface was still not smooth enough for the contact line.

On smooth solid surfaces, Starov et al.[21] and Arjmandi-Tash et al.[22] have shown that equilibrium and hysteresis contact angles can be predicted via disjoining pressure isotherms. Microscopic liquid profile change and a transition region with a critical marked point were proposed to allow the contact line being macroscopically pinned. For the partial-wetting system, they have calculated the hysteresis with estimated parameters and the results were comparable to the experiments. As claimed, the disjoining pressure approach has thus excluded the necessity of surface roughness and resolved the puzzle of the hysteresis on smooth surfaces.[25, 27, 29-30] Now our results have shown that first, the absolutely surface will not have any hysteresis, and secondly, the puzzle of the hysteresis on smooth surfaces[25, 27, 29-30] can be explained since those surfaces were still made of atoms and therefore not smooth enough for the contact line as seen in our simulation and experiments. Just like friction would arise between two atomically smooth solids in contact, the friction would arise between a liquid and an atomically smooth solid.

A straightforward approach for the hysteresis is through the understanding of friction. For real surfaces, the liquid-solid interfacial potential energy can be expressed as,[47]

$$E_{int} = \sum_{n=1}^{N} \varepsilon \cos(\frac{2\pi}{a} x) \tag{3}$$

where *a* is the lattice constant on a solid surface, *n* the sequence number of atoms along the interface, *ε* the intermolecular interaction between solid and liquid molecules, *x* the lateral direction of the solid surface. The periodic potential field can be simplified as shown in Fig. 1 for surface B. The static friction was given by the maximum of the derivative of potential energy with respect to the lateral distance according to the friction theories based upon energy arguments. [47]

$$F_s = \left(-\frac{\partial E_{int}}{\partial x}\right)_{max} = \sum_{n=1}^{N} \frac{2\pi}{a}\varepsilon \qquad (4)$$

As shown in Table 1, the static friction calculated based on Eq. 4 had the same trend of the hysteresis variation in the simulation i.e. greater roughness brought greater static friction and thus the hysteresis. According to Eq. (4), the static friction can also be enlarged by greater liquid-solid adhesion i.e. smaller equilibrium contact angle. The good agreement to the simulation result is also shown in Table 1. In experiments, observation of hysteresis increasing with wettability has also been reported. Extrand [25] measured the contact angle hysteresis of water and organic droplets on polychlorotrifluoroethylene (PCTFE) and polyethylene terephthalate (PET). These two substrates had almost the same roughness while had different wettability. The experimental results showed that hysteresis increased with wettability for all tested liquids.

3.2. Moving contact line

On the absolutely smooth surface A, we have seen that the droplet started to slide on surface A with an acceleration when the lateral external force was applied. The angles were fixed at equilibrium angle despite the movement of the contact line which meant the varying dynamic contact angle was not generated. In contrast, the deviation from the equilibrium angle would be seen on surface B as long as the external force was applied.

Here we would address about an important mesoscopic feature for moving contact line. Recently Chen *et al.* [37] clarified the nanoscopic morphology of the advancing contact line using an AFM. The key feature was a shoe-tip-like convex nanobending region within 20 nm of the substrate. The profile of the nanobending varied with the moving speed and the bending vanished when the speed was zero. The nanobending was believed to be an important mesoscopic structure since it served as the link between molecular and macroscopic domains.

As shown in Fig. 3(a), the convex nanobending was for the first time reproduced in our ultra-large-scale molecular dynamics simulation of droplet spreading. The bending according to the experiment was the deviation of the local profile from the extrapolation of the bulk profile when it comes to the substrate. The microscopic contact angle, $\theta_m$, was formed at the root of the nanobending. The extrapolation of the droplet bulk profile intersected with the solid substrate at point Π and formed the apparent contact angle, $\theta_a$. The simulated nanobending height as shown in Fig. 3 was around 18 nm, which was highly consistent with Chen's experimental results on atomically smooth surfaces, around 20 nm.[37] Note in previous attempts, Blake *et al.* [48-49] and Lukyanov *et al.* [50] didn't observe the nanobending in their simulations. The reason was as Lukyanov *et al.* speculated, the simulation domains were not big enough. As shown in the experiment and in the current simulation, the bending itself has a height about 20 nm for atomically smooth surfaces. To capture the nanobending one has to have a bigger domain since the bending appears as a contrast to the bulk liquid profile. Moreover, extra gravity was applied in the current simulation to have the same Bond number as the droplet condition in the experiment.[37] More information about the nanobending simulation refer to the Supporting Information.

In sharp contrast, the convex nanobending vanished on absolutely smooth substrate A as shown in Fig. 3(b). Note both surfaces B and A had the same equilibrium contact angle, 56°. The only reason that substrate A didn't have the nanobending was that there was no roughness and thus no friction on it. Substrate A had no nanobending no matter how the equilibrium contact angle varied. On surface B, the local friction held the contact line from moving forward and thus produce the convex bending. As shown in Fig. 3(c), the intersect of the bulk profile extrapolation on the solid, point Π, was spreading faster than the contact line, the point I. The friction at the contact line had stronger holding on the local region than that in the bulk, which produced the convex bending. By introducing greater energy barrier for the contact line, as shown in Fig. 3(d) with rougher surface C, the increase of the roughness brought greater dynamic friction, dragged the contact line to be slower, and thus made the greater scale of the nanobending. The comparison of the snapshots of the nanobending on different solids refers to the Supporting Information.

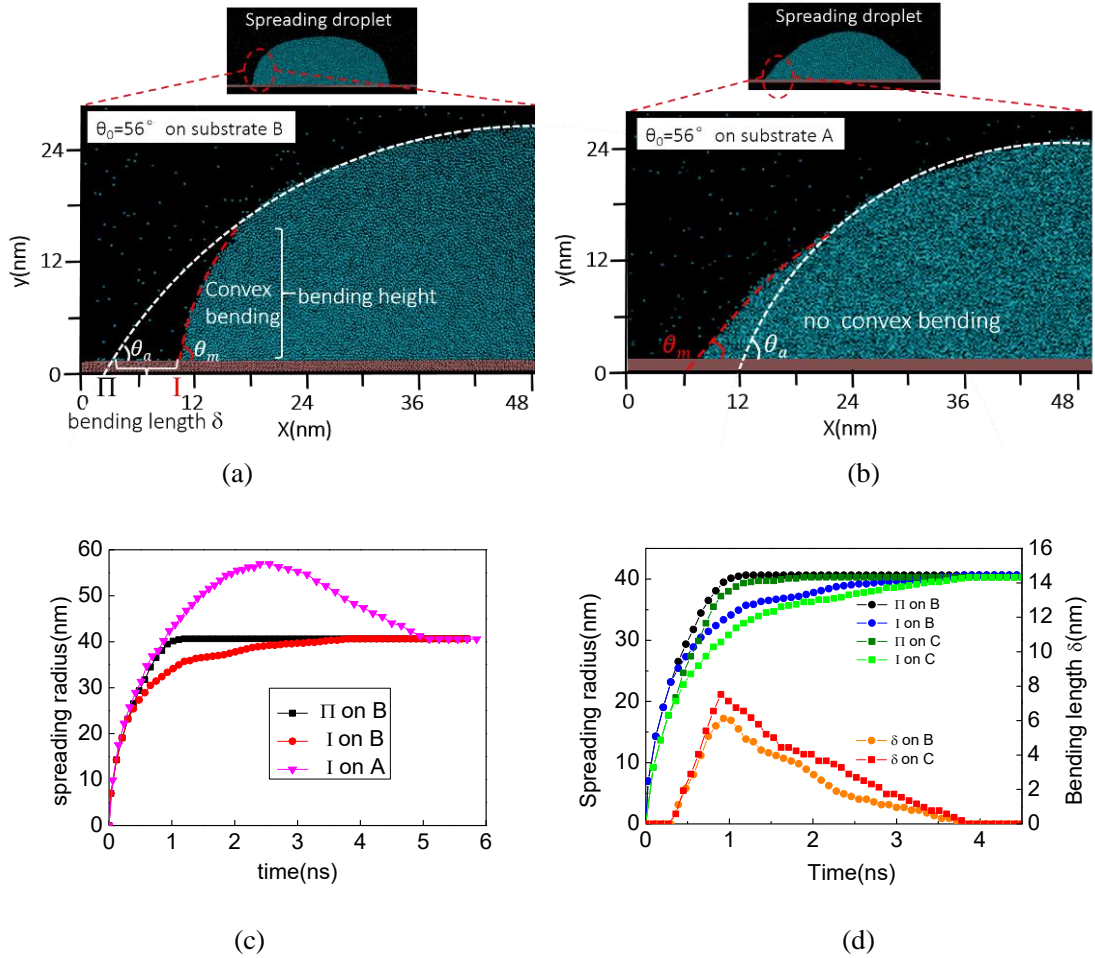

Fig. 3. The convex nanobending at the advancing contact line in the ultra-large-scale molecular dynamic simulation: (a) The nanobending reproduced on atomically smooth surface B, at contact line speed 2 m/s. The bulk extrapolation profile intersected with the substrate at point Π and formed the apparent contact angle, $\theta_a$. The length between point I and Π was defined as the bending length $\delta$. (b) The nanobending vanished on the absolutely smooth surface A. The equilibrium contact angles on the two surfaces were the same, 56°. (c) The spreading of the point I and II on substrate A and B. The spreading was much faster on the surface A than on B. (d) The spreading of the point I and II and the nanobending scale on substrate B and C. Surface C has higher roughness thus slower contact line spreading and greater nanobending than B.

We further conducted AFM experiments to confirm the relationship between the nanobending scale and the roughness. The details refer to Fig. s5 in the Supporting Information. As an example shown in Fig. 4, PEG400 contact line profiles on silicon wafer and quartz were obtained by the tapping-mode AFM. The equilibrium contact angles were about the same, 30°, on both substrates, while the RMS roughness of the quartz was about thirteen times of the silicon wafer. As the result showed, the height of the convex nanobending on the quartz was about 3.8 times of that on the silicon wafer. PPG2000 and glycerol had similar results.

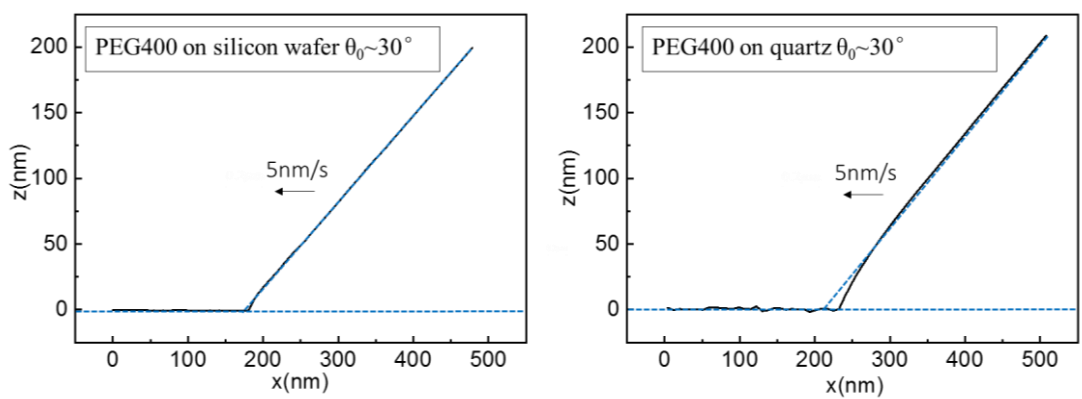

Fig. 4. AFM measurements showing greater nanobending on rougher substrates. The measured RMS roughness of silicon wafer and quartz were 0.42 nm and 5.41 nm respectively.

**4. Conclusions and outlooks**

This work for the first time employed absolutely smooth solid surfaces in ultra-large-scale molecular dynamic simulations to decouple the tangled factors behind the wetting phenomena. The contact angle hysteresis, the variation of the dynamic angle, and the convex nanobending at the advancing contact line all vanished on the absolutely smooth surfaces. In sharp contrast, the solids that made of atoms, even at atomic smoothness, would bring significant deviation from the equilibrium angle and the convex nanobending at advancing contact line. The hysteresis and moving contact line problems were therefore unified to originate from the static and dynamic friction of the contact line on the solid surface. Greater roughness meant greater energy barrier thus induced greater hysteresis and greater scale of convex nanobending. On the other hand, the results have proven that even the minimum i.e. atomic roughness was great enough to be effective in pinning and distorting the contact line as the result of the local friction. The 3D observation using HIM revealed the ubiquitous and non-negligible distortion of atomic roughness on the contact line.

The current disjoining pressure approach may be confusing since it essentially ignores the lateral energy barrier on the surface. A better understanding of static and dynamic friction at the contact line could be the right direction. The force distribution at the moving contact line could be evaluated on the basis of a general friction law for liquid flow at solid surfaces as proposed in the literature[50] however the situation at contact line is still far more complicated than a merely liquid-solid interface.

**ACKNOWLEDGMENTS**

This work was supported by the National Natural Science Foundation of China, No. 51622601 and National Key R&D Program of China, No. 2016YFB0600605.